\documentstyle[editedvolume]{crckapb} 

\begin{opening}
\title{LUMINOUS IR GALAXIES IN A MERGER SEQUENCE: BIMA CO IMAGING}

\author{Y. GAO}
\author{R.A. GRUENDL}
\institute{University of Illinois, Dept. of Astronomy\\
1002 W. Green St., Urbana, IL 61801, USA}
\author{C.Y. HWANG}
\author{K.Y. LO}
\institute{Academia Sinica, IAA\\
P.O. Box 1-87, Nankang, Taipei, Taiwan 11529, ROC}

\end{opening}

\runningtitle{LUMINOUS IR GALAXIES IN A MERGER SEQUENCE}

\begin{document}


\def \eg           {{e.g.}}
\def \ni           {{\noindent}}
\def \etal         {{et~al. }}
\def\approxgt{\lower.2em\hbox{$\buildrel > \over \sim$}}
\def\approxlt{\lower.2em\hbox{$\buildrel < \over \sim$}}
\def \ls{\hbox{L$_{\odot}$}}
\def \ms{\hbox{M$_{\odot}$}}
\def \kms{\hbox{km$\,$s$^{-1}$}}
 
\section{Introduction}

The infrared power output in luminous infrared galaxies (LIGs, 
$L_{\rm IR} \approxgt 10^{11}\ls$, $H_0=75 \kms$ Mpc$^{-1}$) 
can approach the bolometric luminosity of quasars and can be
provided by either starbursts or dust-enshouded QSOs, or both.
Most LIGs appear to comprise of mergers of gas-rich galaxies.
So, intense bursts of star formation apparently result from 
interaction and merging of galaxies, but the exact physical
processes involved in collecting the large amount of gas
involved and in initiating the starbursts are not well understood.  

In order to trace observationally the conditions in the interstellar 
medium (ISM) that lead to starbursts, we have used the newly expanded 
Berkeley-Illinois-Maryland Association (BIMA) millimeter-wave array
to map the molecular 
ISM in a sample of LIGs chosen to represent different phases of the 
interacting/merging process. Most importantly, a few LIGs in the sample
might be in a ``pre-starburst'' phase, so they provide an ideal 
laboratory for studying the conditions leading to starbursts 
(Lo, Gao \& Gruendl 1997).  Our emphasis is on the widely separated 
LIGs at early/intermediate stages of interaction for this reason, and
also for complementary studies to previous CO imaging studies that
have concentrated on relatively
advanced merger systems (Scoville \etal 1991; Downes \& Solomon 1997)
in which the ISM has already been highly 
disrupted by the interaction and starbursts. 

Our goal is to sample statistically the evolution of physical conditions
of the molecular material in LIGs, and to delineate observationally the
development of starburst activities along the merger sequence, using
radio continuum, IR and other observations.

\section{Interacting Luminous Infrared Galaxies}

The observed decrease of $L_{\rm CO}$ with decreasing projected 
nuclear separation of a sample of interacting/merging LIGs can be
interpreted as gas depletion resulting from the merger-enhanced 
starbursts (Gao \& Solomon 1998).  

Unlike the morphologically selected ``Toomre sequence'', the LIGs
in our sample have a comparable gas content so that the galaxies
in the early stage are likely to have the gas reservoir to reach
the ultraluminous starburst phase.  Therefore, they are likely to be
a better sample for the statistical study of interaction/merger induced
starbursts leading to ultra-luminous IR galaxies.
 
\section{CO Imaging of a Merger Sequence}

We present our CO images in Fig.~1 and the gas properties along a merger
sequence in Table~1.

In Table~1, we list the
sample galaxies roughly in a merger sequence 
including a few late stage mergers from literature (marked with *), thus a 
study of the molecular gas properties
at various phases of the merging process can be performed. These are the 
much needed observations to test those sophisticated simulations (\eg,
Barnes 1998).

\begin{table}[htb]
\begin{center}
\caption{Luminous Infrared Galaxies in a Merger Sequence.}
\begin{tabular}{lcrccccc}
\hline 
Source &\multicolumn{1}{c}{R$_{\rm Sep}^1$} & L$_{\rm IR}$ & 
\multicolumn{1}{c}{M(H$_2$)$^2$} & Beam &
\multicolumn{1}{c}{CO$^3$} &
\multicolumn{1}{c}{$\Sigma_{\rm H_2}^4$} & \multicolumn{1}{c}{L$_{\rm IR}$/M(H$_2$)}\\
 & kpc & 10$^{11}$\ls & 10$^{10}$\ms & $''$
\ \ kpc & & 10$^3$\ms pc$^{-2}$ & \ls/\ms \\
\hline
ARP302& 25.8 & 4.1 & 8.0 & 6.0 \ \ 3.7 &u+e&  0.4 & 5.0  \\
N6670 & 14.6 & 3.8 & 5.5 & 4.8 \ \ 2.7 &u+e&  0.3 & 6.9   \\
U2369 & 13.1 & 3.9 & 3.4 & 6.1 \ \ 3.6 &u+e&  1.2 & 11.5 \\
ARP55 & 10.7 & 4.7 & 5.8 & 4.4 \ \ 3.1 &u+e&  1.3 & 8.1  \\
VV114*&  6.0 & 4.2 & 5.1 & 3.7 \ \ 1.4 & e &  3.3 & 8.2 \\
N5256 &  4.8 & 3.1 & 2.7 & 4.6 \ \ 2.3 &u+e& $>$1.3 & 11.5 \\
MRK848&  4.7 & 7.2 & 3.4 & 3.9 \ \ 2.9 &u+e& $>$1.0 & 21.2 \\
ARP299*& 4.5 & 6.4 & 1.4 & 2.3 \ \ 0.4 & e & 11.3 & 45.0 \\
N6090 &  3.5 & 3.0 & 2.4 & 1.9 \ \ 1.0 & e &  3.0 & 12.5 \\
N1614 &  2.0 & 4.1 & 1.6 & 5.8 \ \ 1.7 & u & $>$1.5 & 25.6 \\
MRK273*& 1.0 & 13.2& 2.7 & 2.3 \ \ 1.5 &u+e&  5.5 & 47.6 \\
ARP220*& 0.4 & 15.0& 3.0 & 1.0 \ \ 0.4 & e &  80 & 50.0 \\
MRK231*& 0.0 & 30.4& 3.5 & 0.9 \ \ 0.7 &u+e&  30 & 77.2 \\
\hline
\end{tabular}
\end{center}
1. Projected separation between the two galaxy nuclei.
2. =4.78 L$_{\rm CO}$ K\kms~pc$^2$, the single-dish CO luminosity.
3. CO morphology, u$\equiv$ unresolved peak; e$\equiv$ extended structures resovled by the beam.
\end{table}

Clearly, a sequence of merging
is observed in the molecular gas traced by CO:

\ni (1). The morphology of the molecular gas changes from weakly disturbed 
and separate gas disks to the disturbed or merged-common-envelope gas disks 
and finally to a single common gas disk for the double nuclei of 
the two galaxies.
 
\ni (2). The spatial CO extent drops from $\sim$ 20 kpc for the
early mergers to a few kpc for the intermediate 
 mergers. Advanced mergers have typical
nuclear CO concentration $\approxlt$ 1 kpc.
 
\ni (3). The corrected face-on central gas surface density 
increases from a few times
10$^2$ \ms pc$^{-2}$ to $> 10^3$ \ms pc$^{-2}$ in our sample.
However, advanced ultraluminous mergers typically
have $> 10^4$ \ms pc$^{-2}$. A rapid increase of the nuclear gas 
surface density is evident along the sequence.
 
\ni (4). The ${\rm L_{IR}/M(H_2)}$ ratio (a measure of 
star formation efficiency, SFE) 
increases by roughly several times from the early mergers to the 
intermediate/advanced mergers. We can estimate 
the central SFE ratio, by scaling the far-IR luminosity and extent 
with those of the radio continuum emission or the
mid-IR emission (Hwang \etal 1998), 
which tends to increase more drastically than 
the global SFE along the sequence.
 
\ni (5). We found that early stage mergers or pre-mergers 
like Arp~302 and NGC~6670 appear
to have much smaller SFE throughout the entire interacting/merging disks, 
comparable to that of GMCs in the Milky Way disk. This strongly 
suggests that LIGs in the pre-merging stage
are in a \underbar{pre-starburst} phase (Lo, Gao \& Gruendl 1997). 
  
\ni (6). Starburst phase as indicated by large SFE's 
seems to start once the molecular gas disks
begin merging and the ultraluminous 
starburst phase seems to be occcuring only after the molecular
gas disks merge into a common disk in late stage mergers.

\ni (7). Gas seems to respond faster than the steller component 
when merging advances to late stage (i.e., the gas disks merge 
faster than the steller disks).

\section{References}

Barnes, J.E. 1998, this volume \\
Downes, D., \& Solomon, P.M. 1997, ApJ, submitted \\
Gao, Y., \& Solomom, P.M. 1998, ApJ, submitted \\
Hwang, C.Y. \etal 1998, this volume \\
Lo, K.Y., Gao, Y., \& Gruendl, R.A. 1997, ApJ, 475, L103 \\
Scoville, N.Z. \etal 1991, ApJ, 370, 158 \\

\begin{figure}
\includegraphics{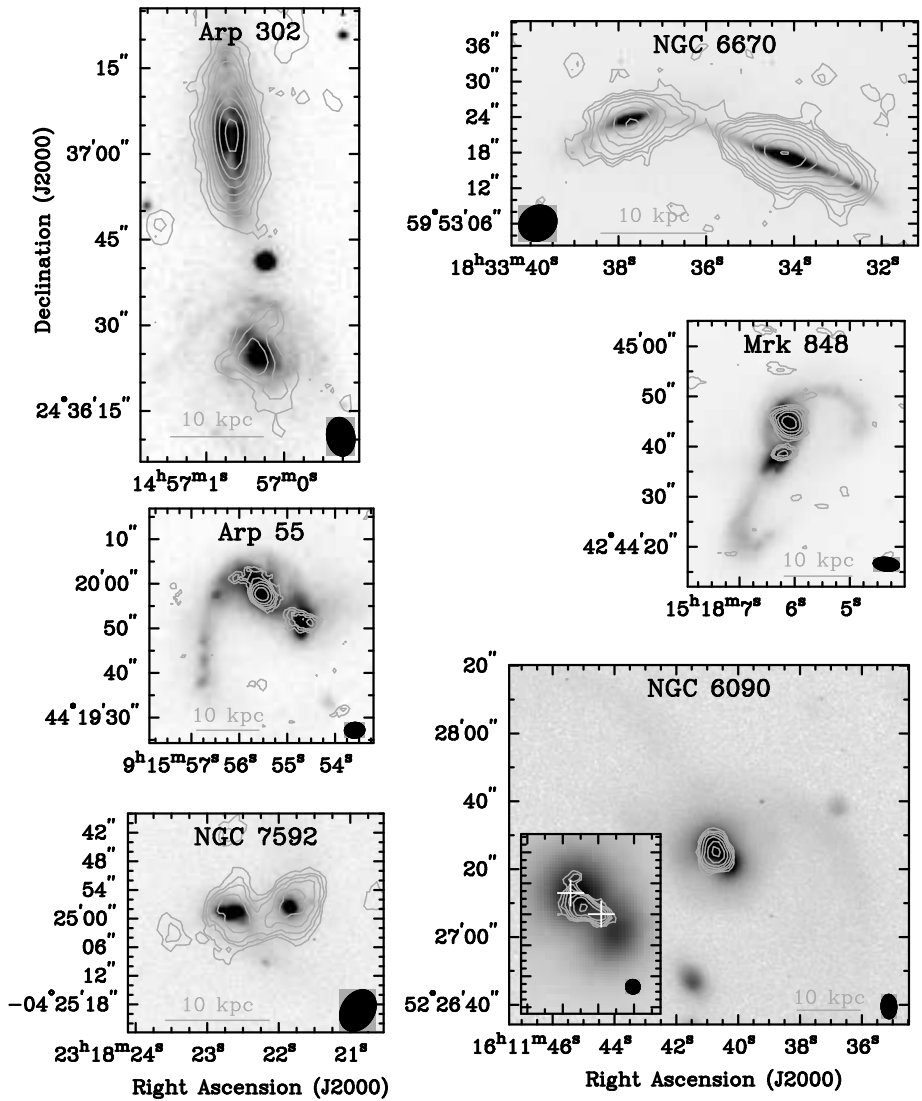}
\vspace{16cm}  
\caption{A merger sequence leading to ultraluminous starburst (CO contours
overlaid on optical images). The insert in N6090 shows higher resolution
($\sim 2''$) CO contours with two radio continuum peaks marked.}
\end{figure}

\end{document}